\documentclass[%
reprint,
 showpacs,showkeys,
 amsmath,amssymb,
 aps,
]{revtex4-1}
\usepackage{amsmath}
\usepackage{cases}
\usepackage{amssymb}
\usepackage{CJK}
\usepackage{graphicx}
\usepackage{dcolumn}
\usepackage{bm}
\usepackage{color}
\usepackage[pagewise]{lineno}

\begin{document}
\begin{CJK*}{GBK}{} 

\preprint{APS/123-QED}

\title{Predictions for cross sections of light proton-rich isotopes in $^{40}$Ca + $^{9}$Be reaction
}
\author{Yi-Dan SONG $^{1}$}
\author{Hui-Ling WEI $^{1}$}
\author{Chun-Wang MA $^{1,2}$}
\thanks{Email address: machunwang@126.com}

\affiliation{$^{1}$
Institute of Particle and Nuclear Physics, Henan Normal University, \textit{Xinxiang 453007}, China \\
$^{2}$
Shanghai Institute of Applied Physics, Chinese Academy of Sciences, \textit{Shanghai 201800}, China
}



\date{\today}

\begin{abstract}
The cross sections for the $Z=$ 10 -- 19 with isotopes $T_{z}=-3/2$ to $-5$ in the 140$A$ MeV $^{40}$Ca + $^{9}$Be projectile fragmentation reaction have been predicted. An empirical formula based on the correlation between the cross section and average binding energy of isotope has been adopted to predict the cross section. The binding energies in the AME16, WS4, and the theoretical prediction by the spherical relativistic continuum Hartree$-$Bogoliubov theory have been used. Meanwhile, the {\sc fracs} parametrization and the modified statistical abrasion-ablation model are also used to predict the cross sections for the proton-rich isotopes. The predicted cross sections for the $T_{z}=-$3 isotopes are close to $10^{-10}$ mb, which hopefully can be studied in experiment. In addition, based on the predicted cross sections, the $Z=$ 14 is suggested to be a new magic number in the light proton-rich nuclei with $T_{z} \leq-$3/2, of which the phenomenon is much more evident than it is from the average binding energy per nucleon.
\end{abstract}

\maketitle
\end{CJK*}


\section{introduction}
More than 8,000 nuclides have been predicted to be bound and have lifetimes longer than 1 $\mu s$ \cite{IsoPred18MJ,AME16,BNNmass,NWang14PLB}. Based on the improved ability of new radioactive nuclear beam (RNB) facilities, for example, the RIKEN-BigRIPS \cite{BigRips}, HIRFL-RIBLL/RIBLL2 \cite{RIBLL2SciBul,RIBLL2}, FRIB-ARIS \cite{FRIB} etc., isotopes near the drip lines can be studied experimentally. Besides the great interests in the neutron-rich isotopes, the proton-rich (also refers to neutron-deficient) isotopes have also attracted great attention from experimentalists. The reported experimental results have illustrated exotic phenomena in proton-rich isotopes, such as the proton halo and core deformation \cite{PHalo,CoreDef}, ($\beta-$delayed) one or two protons emissions \cite{ProtEmisCa,ProtEmisAl,ProtEmisSi,ProtEmisMg,Kr78Exp,Kr67Emis}. Moreover, the researches for new isotopes near the proton-drip line are also motivated by new physics related to the very isospin asymmetry of new magic number, new single particle structure, short-range correlation, etc. These studies explore the properties of isotopes near the proton-drip line, which are important both for nuclear physics and astrophysics.

In designing an experiment, it is important to predict the cross sections for proton-rich isotopes. Theoretical models include the molecular model \cite{Mocko08,Huang10IYR,Ma15CPL,BianBA}, the modified statistical abrasion-ablation (SAA) model \cite{MaCW09PRC}, etc. The readers are referred to a recent review \cite{PPNP18Ma} for more information. For the proton-rich isotopes the theoretical predictions are less precise compared to the neutron-rich ones due to the limited data reported. A parametrization proposed by B. Mei named as {\sc fracs} \cite{FRACS} improved the quality of the {\sc epax3} parametrization \cite{EPAX3} by incorporating the dependence of isotopic cross section on the incident energy. Some empirical formula can also be applied to predict the isotopic cross sections \cite{MaCW17JPG,BvsCRSCu}. For example, an empirical correlation between the binding energy and isotopic cross section has been used to predict the cross section and the binding energy of neutron-rich copper isotopes with high precise \cite{BvsCRSCu}. In this article, we will report the predicted cross sections for some proton-rich isotopes in the $^{40}$Ca + $^{9}$Be projectile fragmentation reaction.

\section{Methods}
\label{method}
Based on the statistical thermodynamics theory, the isotopic yields are related to the free energy of fragments \cite{PPNP18Ma}. Different methods have been suggested to describe the free energy of fragment. For examples, a temperature-dependent parametrization of binding energy \cite{Huang10IYR}, the decaying modifications including the pairing effects and particle emissions \cite{BvsCRSCu,Ma12T,Ma13T,Ma15T}. Besides these works, empirical formula for isotopic cross section has also been proposed by investigating experimental data. A simple correlation between the isotopic cross section and the binding energy has been proposed by Tsang \textit{et al}. \cite{BvsCRSCu} For the neutron-rich fragment, the isotopic cross section is found to depend exponentially on its average binding energy,
\begin{equation}\label{EMPFor1}
\sigma=C \exp[(<B'>-8)/\tau],
\end{equation}
where $C$ and $\tau$ are free parameters. $<B'>$ is the average binding energy, which is defined as,
\begin{equation}
<B'>=(B-\epsilon_{p})/A,
\end{equation}
in which $\epsilon_{p}$ is related to the pairing energy,
\begin{equation}
\epsilon_{p}=0.5\left[(-1)^{N}+(-1)^{Z}\right]\epsilon_{0} A^{-3/4}.
\end{equation}
$\epsilon_{p}$ is introduced to minimize the odd-even staggering in isotopic cross section distribution, and $\epsilon_{0}=$ 30 MeV is adopted according to Ref. \cite{BvsCRSCu}. Using this correlation, the binding energies of the neutron-rich $^{76-79}$Cu isotopes have been extracted. Meanwhile, the cross sections for the near drip-line isotopes ($^{39}$Na and $^{40}$Mg) have also been predicted \cite{BvsCRSCu}. A similar work predicted the binding energies of proton-rich isotopes in the 345$A$ MeV $^{78}$Kr + $^9$Be reaction \cite{YDS18HNU}. The cross sections for fragments in the 140$A$ MeV $^{40}$Ca + $^{9}$Be projectile fragmentation reaction have been measured by Mocko \textit{et al} \cite{Mocko06,MockoTh}, in which the proton-rich isotopes with $T_{z} \geq -$1 are reported. Based on the measured cross sections for proton-rich isotopes, the correlation in Eq. (\ref{EMPFor1}) will be applied to the proton-rich isotopes in the 140$A$ MeV $^{40}$Ca + $^{9}$Be reaction. For the proton-rich isotopes, we do not intend to modify the empirical formula since the mass formula can well describe most of the isotopes. The cross sections for isotopes with $T_{z}$ from $-$3/2 to $-$5 will be predicted using Eq. (\ref{EMPFor1}), as well as the {\sc fracs} parametrization \cite{FRACS} and the modified SAA model \cite{MaCW09PRC,Ma15CPL}.

\begin{figure}[htbp]
\includegraphics
[width=8.6cm]{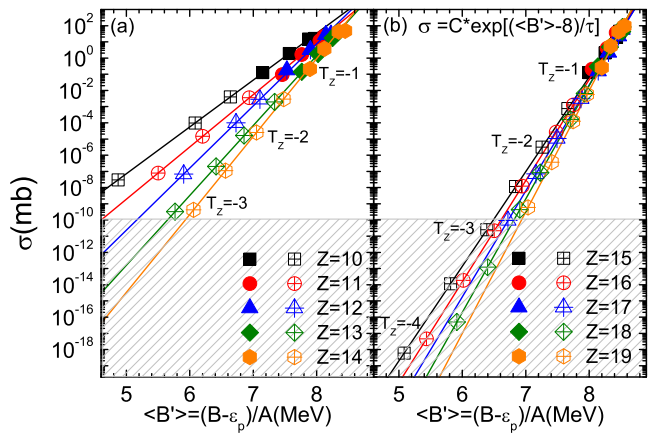}
\caption{\label{IsoCrsZ1019} (Color online) Correlation between $\sigma$ and $<B'>$ of $Z=$ 10--19 isotopes in the 140$A$ MeV $^{40}$Ca + $^{9}$Be reaction. The solid symbols denote the measured $\sigma$ taken from Ref. \cite{Mocko06}, and the lines denote the fitting results to the correlation for the measured results of each isotopic distribution using Eq. (\ref{EMPFor1}). The open symbols are the predicted $\sigma$ using the fitting functions by adopting the binding energies in AME16 \cite{AME16} or the RCHB theory \cite{IsoPred18MJ} (see text).
}
\end{figure}

\section{results and discussion}
\label{RAD}
\begin{figure}[htbp]
\includegraphics
[width=8.cm]{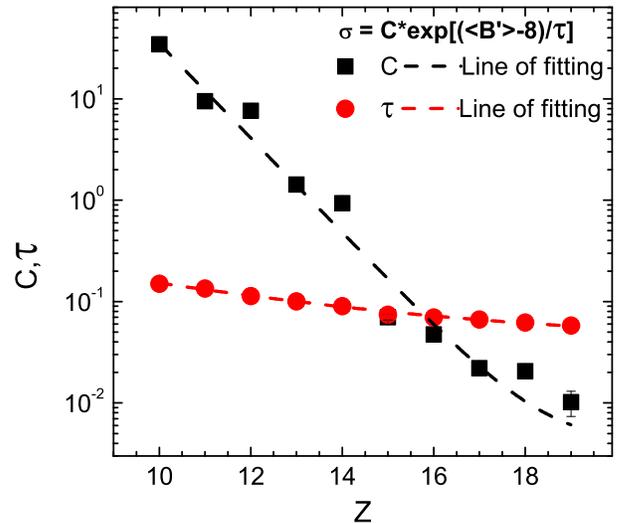}
\caption{\label{CKvalue} (Color online) The values of $C$ and $\tau$ obtained from the measured isotopes of $Z=$ 10--19 in the 140$A$ MeV $^{40}$Ca + $^{9}$Be reaction.
}
\end{figure}

In Fig. \ref{IsoCrsZ1019}, the correlation between the measured cross sections (solid symbols) and $<B'>$ for the $Z=$ 10 (Ne) to 19 (K) isotopes, with $T_{z}=$ $-$1 to 1/2, are plotted. For the measured isotopes, the experimental binding energies are taken from AME16 \cite{AME16}. The correlations between $\sigma$ and $<B'>$ can be well fitted by Eq. (\ref{EMPFor1}). The values of $C$ and $\tau$ for each isotopic distribution have been obtained, which are plotted in Fig. \ref{CKvalue}. It is seen that $C$ decreases with the increasing $Z$. Though $\tau$ also decreases with the increasing $Z$, the dependence is much weaker than $C$. The fitting result to the $C-Z$ correlation reads $C= 0.00387+1.62\times 10^{6}\exp(-1.07395\cdot Z)$, and the fitting result to the $\tau-Z$ correlation is $\tau= 0.04+0.909\exp(-0.2089\cdot Z)$, which both can be well fitted by the exponential function. The dependence of $C$ and $\tau$ on $Z$ will be discussed later.

With the values of $C$ and $\tau$ known, one can predict the cross sections for the more proton-rich isotopes in the 140$A$ MeV $^{40}$Ca + $^{9}$Be reaction using Eq. (\ref{EMPFor1}). In the prediction, if the experimental binding energy of an isotope exists in AME16 \cite{AME16}, the experimental one is used. If the experimental binding energy is absence, the prediction in a recent work by the spherical relativistic continuum Hartree$-$Bogoliubov (RCHB) theory with the relativistic density functional PC$-$PK1 \cite{IsoPred18MJ} will be adopted. In Fig. \ref{IsoCrsZ1019}, one can see the predicted cross sections for the isotopes with $T_{z}<-1$ (open symbols). As $T_{z}$ decreases, the predicted $\sigma$ for isotopes drop fast.
The cross sections for isotopes measured in experiments touched $\sigma\sim 10^{-10}$ mb \cite{BlankGe70Exp,Stolz}. If we consider $\sigma\sim 10^{-10}$ mb as the lower limitation for isotopes, the isotopes with $T_{z}\geq -3$ are possible candidates to be studied in present RNB facilities. With the enhanced beam intensity for rare isotopes, it is suggested that in the new RNB facilities more opportunity will be opened for the very proton-rich isotopes with $T_{z}<-$3, and it is also possible to fix the locations of proton-drip line for the $Z =$ 10--19 isotopes.

\begin{figure}[htbp]
\includegraphics
[width=8.6cm]{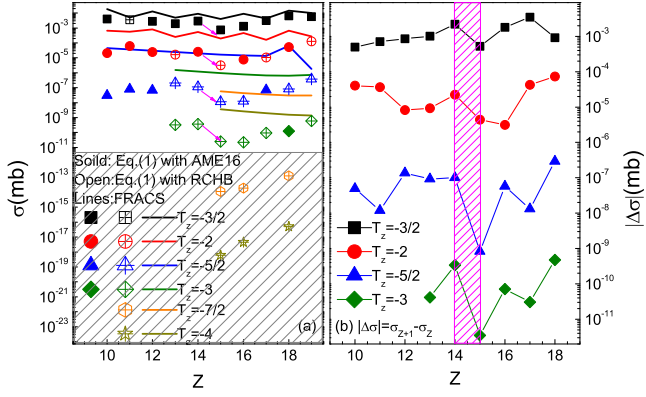}
\caption{\label{IsoPred} (Color online) (a)The predicted cross sections for the $Z=$ 10--19 isotopes with $T_{z}=-3/2$ to $-4$ in the 140$A$ MeV $^{40}$Ca + $^{9}$Be reaction. The symbols denote the predictions using Eq. (\ref{EMPFor1}), with the solid and open ones using the measured data in AME16 \cite{AME16} and the predictions by RCHB \cite{IsoPred18MJ}, respectively. The lines denote the predictions using the {\sc fracs} parametrization \cite{FRACS}. The arrows indicate the jumps from the $Z=14$ to 15 nuclei in each $T_{z}$ chains (see text). (b) The gap ($|\Delta\sigma|$) between $\sigma$ of $Z+1$ and $Z$ isotopes $|\Delta\sigma|=\sigma_{Z+1}-\sigma_{Z}$.
}
\end{figure}

Besides the predictions using the binding energies in AME16 and RCHB, more results are obtained using the WS4 binding energy \cite{NWang14PLB}, the {\sc fracs} parametrization \cite{FRACS}, and the SAA model \cite{MaCW09PRC}. In Fig. \ref{IsoPred}(a), the predicted $\sigma$ for isotopes with $T_{z}$ from $-$4 to $-$3/2 by Eq. (\ref{EMPFor1}) and the {\sc fracs} parametrization are plotted. It can be seen that the predicted $\sigma$ for each $T_{z}$ chain almost forms a plateau. The predicted $\sigma$ for the $T_{z}=-$3/2 nuclei by {\sc fracs} are close to the results of Eq. (\ref{EMPFor1}). While for the isotopes from $T_{z}=$-2 to -4, the predicted $\sigma$ by {\sc fracs} are larger than those by Eq. (\ref{EMPFor1}). Considering the predicted results by Eq. (\ref{EMPFor1}), $\sigma$ for the $T_{z}=-$3 nuclei are close to $\sim 10^{-10}$mb, while $\sigma$ for the $T_{z}=-$7/2 nuclei drop to $\sim 10^{-14}$ mb and are much lower than the $T_{z}=-$3 nuclei. Taking $10^{-10}$mb as a lower limitation in experiment, it is suggested that the $T_{z}\geq-$3 isotopes can be considered in future experiments, while the $T_{z} <-$3 isotopes maybe much more difficult to be produced in present RNB facilities.

The predicted $\sigma$ for the proton-rich isotopes in the 140$A$ MeV $^{40}$Ca + $^{9}$Be reaction are listed in Table \ref{PredCrS}. One can see that the predictions using Eq. (\ref{EMPFor1}) but with different binding energies from AME16 \cite{AME16}, WS4 \cite{NWang14PLB} and RCHB \cite{IsoPred18MJ} are close for the $T_{z} \geq -$2 isotopes. For some of the $T_{z} = $ --5/2 and --3 isotopes, the predicted $\sigma$ using the different binding energies show more obvious difference. The {\sc fracs} and SAA model predictions in general are larger than results using Eq. (\ref{EMPFor1}). The reason is that in the {\sc fracs} the cross sections for the proton-rich isotopes are not delicately adjusted due to the lack of experimental data. While in the SAA model, the secondary decay of hot fragments significantly depend on the separation energy calculated from the mass formula, which has a low accuracy for the proton-rich isotope.

\begin{table}[thbp]
\caption{The cross sections for the $Z=$ 10 -- 19 isotopes in the 140$A$ MeV $^{40}$Ca + $^{9}$Be reaction predicted using Eq. (\ref{EMPFor1}) but with different binding energies, the {\sc fracs} parametrization \cite{FRACS} and the SAA model \cite{MaCW09PRC}.}
\label{PredCrS}
\centering
\begin{tabular}{p{25pt}p{20pt}p{32pt} p{32pt}p{32pt}p{1pt} p{32pt}p{32pt} }
  \hline
  \hline
  $^{A}$Z   &T$_{Z}$ &             &Eq.(\ref{EMPFor1}) &          &&Theories      &\\
                     \cline{3-5}                                   \cline{7-8}
            &        &AME16        &WS4                &RCHB      &&{\sc fracs}   &SAA\\
  \hline
  $^{17}$10 &-3/2    &4.0E-3       &--                 &4.0E-3    &&1.9E-2        &1.1E-2 \\
  $^{19}$11 &-3/2    &3.5E-3       &3.3E-3             &3.5E-3    &&5.1E-3        &3.5E-3 \\
  $^{21}$12 &-3/2    &2.8E-3       &3.0E-3             &2.8E-3    &&1.3E-2        &6.7E-3 \\
  $^{23}$13 &-3/2    &1.9E-3       &1.9E-3             &1.9E-3    &&4.9E-3        &7.6E-3 \\
  $^{25}$14 &-3/2    &3.0E-3       &3.0E-3             &3.0E-3    &&8.9E-3        &1.0E-2 \\
  $^{27}$15 &-3/2    &7.6E-4       &7.6E-4             &7.5E-4    &&4.0E-3        &1.1E-2 \\
  $^{29}$16 &-3/2    &1.3E-3       &1.3E-3             &1.3E-3    &&8.8E-3        &2.9E-2 \\
  $^{31}$17 &-3/2    &3.1E-3       &3.1E-3             &3.1E-3    &&4.4E-3        &5.2E-2 \\
  $^{33}$18 &-3/2    &6.6E-3       &6.6E-3             &6.6E-3    &&1.4E-2        &1.0E-1 \\
  $^{35}$19 &-3/2    &5.7E-3       &5.7E-3             &4.1E-3    &&1.0E-2        &1.5E-1 \\
  \hline
  $^{16}$10 &-2      &2.1E-5       &--                 &2.1E-5    &&6.6E-4        &1.3E-8 \\
  $^{18}$11 &-2      &6.1E-5       &--                 &6.1E-5    &&5.7E-4        &4.9E-5 \\
  $^{20}$12 &-2      &2.5E-5       &2.0E-5             &2.4E-5    &&8.1E-4        &5.1E-5 \\
  $^{22}$13 &-2      &3.0E-5       &2.4E-5             &1.6E-5    &&2.6E-4        &8.2E-5 \\
  $^{24}$14 &-2      &2.6E-5       &2.6E-5             &2.6E-5    &&5.4E-4        &5.3E-4 \\
  $^{26}$15 &-2      &5.5E-6       &3.9E-6             &3.1E-6    &&2.0E-4        &1.0E-3 \\
  $^{28}$16 &-2      &7.5E-6       &7.5E-6             &7.5E-6    &&4.6E-4        &4.2E-3 \\
  $^{30}$17 &-2      &2.8E-5       &1.9E-5             &1.1E-5    &&1.6E-4        &6.5E-3 \\
  $^{32}$18 &-2      &5.3E-5       &5.3E-5             &5.3E-5    &&6.5E-4        &6.0E-3 \\
  $^{34}$19 &-2      &1.0E-4       &6.2E-5             &1.3E-4    &&2.7E-4        &1.8E-2 \\
  \hline
  $^{15}$10 &-5/2    &3.0E-8       &--                 &--        &&4.6E-5        &--     \\
  $^{17}$11 &-5/2    &8.0E-8       &--                 &6.5E-7    &&3.7E-5        &--     \\
  $^{19}$12 &-5/2    &6.8E-8       &--                 &6.8E-8    &&3.0E-5        &1.1E-7 \\
  $^{21}$13 &-5/2    &6.7E-8       &5.5E-8             &2.0E-7    &&2.5E-5        &2.7E-7 \\
  $^{23}$14 &-5/2    &1.2E-7       &8.0E-8             &1.1E-7    &&2.0E-5        &5.1E-5 \\
  $^{25}$15 &-5/2    &8.0E-9       &4.3E-9             &1.1E-8    &&1.7E-5        &9.6E-5 \\
  $^{27}$16 &-5/2    &2.0E-8       &1.3E-8             &1.2E-8    &&1.4E-5        &1.3E-4 \\
  $^{29}$17 &-5/2    &7.0E-8       &3.0E-8             &3.8E-8    &&1.3E-5        &3.4E-4 \\
  $^{31}$18 &-5/2    &1.2E-7       &7.4E-8             &8.3E-8    &&6.4E-5        &1.3E-3 \\
  $^{33}$19 &-5/2    &3.6E-7       &1.8E-7             &3.7E-7    &&1.8E-6        &2.5E-3 \\
  \hline
  $^{20}$13 &-3      &--           &--                 &3.3E-10   &&1.6E-6        &--     \\
  $^{22}$14 &-3      &9.5E-11      &7.7E-11            &3.7E-10   &&1.2E-6        &--     \\
  $^{24}$15 &-3      &9.2E-12      &5.0E-12            &2.5E-11   &&9.5E-7        &2.6E-7 \\
  $^{26}$16 &-3      &9.7E-12      &6.1E-12            &2.2E-11   &&7.7E-7        &4.2E-6 \\
  $^{28}$17 &-3      &7.2E-11      &3.0E-11            &9.3E-11   &&6.6E-7        &2.0E-5 \\
  $^{30}$18 &-3      &1.2E-10      &4.6E-11            &1.4E-10   &&6.3E-7        &1.9E-5 \\
  $^{32}$19 &-3      &4.3E-10      &1.4E-10            &6.0E-10   &&7.1E-7        &1.2E-4 \\
  \hline
  $^{23}$15 &-7/2    &--           &6.9E-16            &1.2E-14   &&5.7E-8        &--     \\
  $^{25}$16 &-7/2    &--           &1.8E-15            &1.9E-14   &&4.4E-8        &--     \\
  $^{27}$17 &-7/2    &--           &4.3E-15            &--        &&3.5E-8        &7.3E-7 \\
  $^{29}$18 &-7/2    &--           &9.9E-15            &1.2E-13   &&3.0E-8        &3.3E-6 \\
  $^{31}$19 &-7/2    &--           &2.6E-14            &--        &&3.0E-8        &8.1E-6 \\
  \hline
  $^{22}$15 &-4      &--           &--                 &6.0E-19   &&3.6E-9        &--     \\
  $^{24}$16 &-4      &--           &1.1E-19            &4.6E-18   &&2.6E-9        &--     \\
  $^{26}$17 &-4      &--           &5.0E-19            &--        &&2.0E-9        &--     \\
  $^{28}$18 &-4      &--           &4.2E-19            &5.0E-17   &&1.6E-9        &--     \\
  $^{30}$19 &-4      &--           &2.5E-18            &--        &&1.4E-9        &--     \\
  \hline
  $^{23}$16 &-9/2    &--           &--                 &7.3E-23   &&1.6E-10       &--     \\
  $^{25}$17 &-9/2    &--           &7.1E-24            &--        &&1.2E-10       &--     \\
  $^{27}$18 &-9/2    &--           &6.6E-24            &--        &&8.5E-11       &--     \\
  $^{29}$19 &-9/2    &--           &2.9E-23            &--        &&6.9E-11       &--     \\
  \hline
  $^{28}$19 &-5      &--           &8.6E-29            &--        &&3.7E-12       &--     \\
  \hline\hline
\end{tabular}\\
The binding energies of isotopes are taken from AME16 \cite{AME16}, WS4 \cite{NWang14PLB}, and RCHB \cite {IsoPred18MJ}.
\end{table}

We now discuss the fitted parameters of $C$ and $\tau$. It is known the values of $C$ and $\tau$ depends on the reaction systems. In the canonical ensemble theory within the grand canonical limitations \cite{BvsCRSCu}, the cross section for a fragment with $(Z,N)$ is,
\begin{equation}\label{YCE-GC}
\sigma(N,Z)=cA^{3/2}\exp[(N\mu_n+Z\mu_p-F)/T],
\end{equation}
where $c$ is a parameter relating to the system, $\mu_n$ ($\mu_p$) the chemical potential of neutrons (protons), $F$ the free energy of fragment, and $T$ the temperature. The equation in Eq. (\ref{YCE-GCd1}) can be changed to the following form by introducing $T_{z}$,
\begin{equation}\label{YCE-GCd1}
\sigma(N,Z)=cA^{3/2}\exp\{[2T_{z}\mu_n+Z(\mu_{n}+\mu_p)-F]/T\}.
\end{equation}

Comparing Eq. (\ref{EMPFor1}) with Eq. (\ref{YCE-GCd1}), $C$ is influenced by the asymmetry and free energy of fragment, as well as the asymmetry of system (via chemical potential); and $\tau$ corresponds to $T$. $T$ determined from the intermediate mass fragments shows a slight dependence on mass \cite{MaCW18JPGT}, which accounts for the slight dependence of $\tau$ on $Z$. With the limited data reported, the 140$A$ MeV $^{40}$Ca + $^{9}$Be reaction is selected in this work for the reason that a series of proton-rich isotopes have been measured in high quality. The isotopic cross sections for the $Z <$ 10 are not investigated since the parametrization for binding energy is not good for them (The RCHB does not report the binding energy for them). Besides, for the light particles, they have different kind of production mechanisms to the intermediate mass fragments. One is hoped to check this correlation at a different reactions since the incident energy influences the isotopic distributions, which calls for more experimental data for proton-rich isotopes. It is supposed that the simple correlation should work for the isotopes produced in the projectile fragmentation reactions from above a few tens of $A$ MeV to the relativistic energy range. Based on these facts, it is suggested that more work should be performed to study the proton-rich isotope production experimentally for a full understanding of proton-drip line nuclei due to their importance in nuclear theory, nuclear astrophysics, etc.

\begin{figure}[htbp]
\includegraphics
[width=8.6cm]{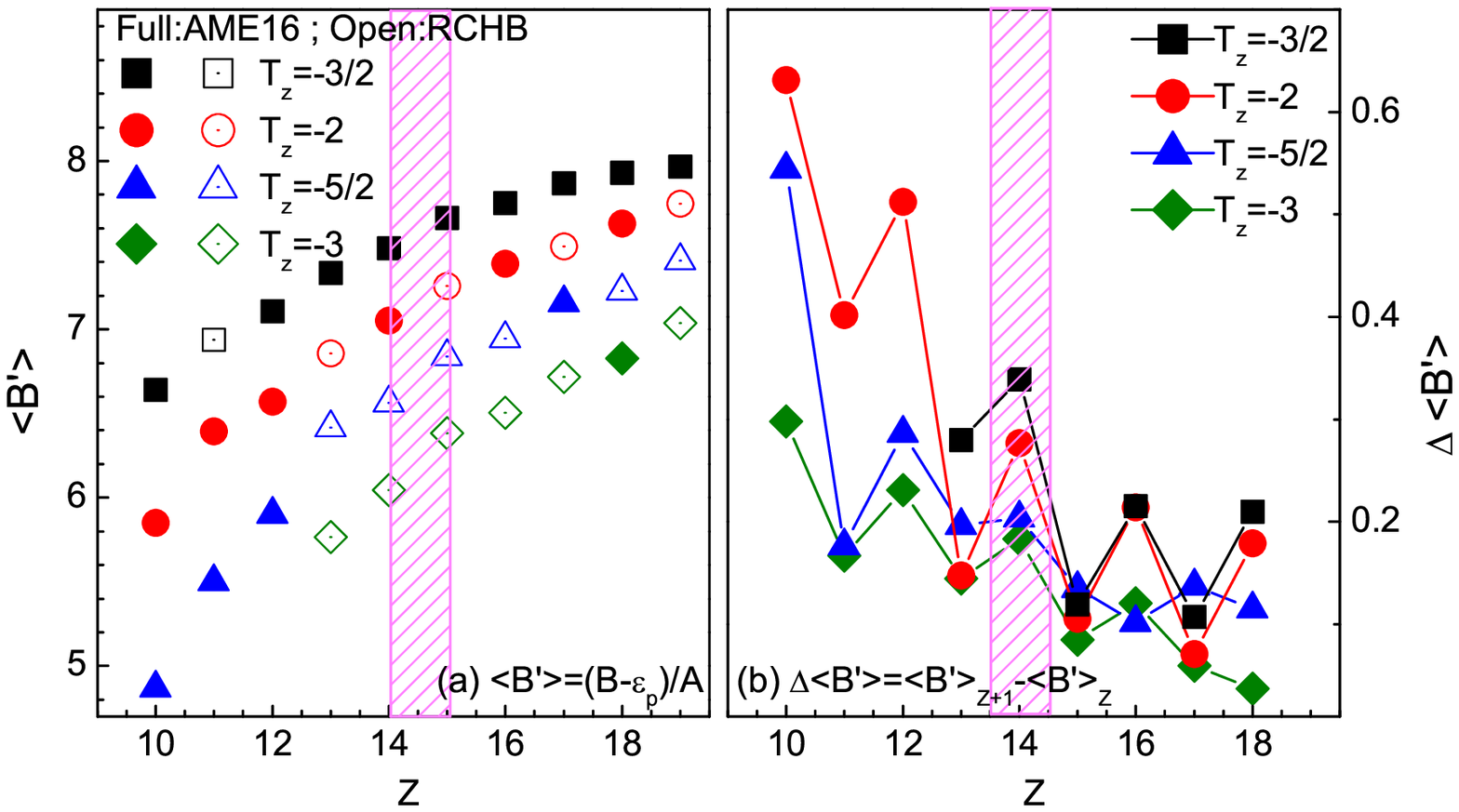}
\caption{\label{BEgap} (Color online) (a) The average binding energy $<B'>$ of the $Z=$ 10--19 isotopes with $T_{z}=-3/2$ to $-3$. (b) The gap ($\Delta<B'>$) between $<B'>$ of the $Z$ and $Z-1$ isotopes $\Delta<B'>=<B'>_{Z+1}-<B'>_{Z}$.
}
\end{figure}

We note that the pairing energy, which originates from the pairing correlation of nucleons, is important for the isotopes with large asymmetry \cite{IsoPred18MJ}. The shell evolution in isotopes will influence the binding energy significantly, and will also results in exotic phenomena in distributions \cite{Ma12CPLasym,MaCW13JPG}. The new magic number of $Z =$ 14 in light nuclei near drip line has been indicated in a prediction within the cluster-core model \cite{Z16Shell}. The trend of the cross section distribution breaks at $Z=14$ for the $T_{z}\leq-$3/2 nuclei (see the arrows in Fig. \ref{IsoPred}(a)). To further show the gap, the quantities of $|\Delta\sigma\equiv\sigma_{Z+1}-\sigma_{Z}|$ are plotted in Fig. \ref{IsoPred}(b). It is clearly shown that $|\Delta\sigma|$ is much lower than the neighboring ones in the $T_{z}$ from --3 to --3/2 nuclide chains. The large gaps between the predicted cross sections for the $Z=$ 14 to 15 nuclei agree with conclusion in Ref. \cite{Z16Shell} that $Z=$ 14 is a new magic number in proton-rich isotopes.

To further check the new $Z=$ 14 magic number, the distributions of $<B'>$ for them are also studied with the combination data in AME16 and RCHB. In Fig. \ref{BEgap}(a), it is seen that the distributions of $<B'>$ go smoothly though some slight pairing effect can be found. In Fig. \ref{BEgap}(b), the gaps between $<B'>$ of the neighboring nuclei, $\Delta<B'>\equiv<B'>_{Z+1}-<B'>_{Z}$, are also plotted. $\Delta<B'>$ does not show a clear evidence of the enhancement of new (sub)shell. This maybe due to that the experimental and theoretical RCHB results are used in a mixed manner, which is difficult to show the shell closure phenomenon. It can be concluded that the cross section enhances the signature of new shell closure for average binding energy. And it is also required that systematic experimental measurements should be performed for the binding energy of these nuclei to verify if there is new magic number at $Z=$ 14.

Furthermore, the cross sections and the information uncertainties of neutron-rich fragments have been found to illustrate an isobaric scaling phenomenon in neutron-rich systems \cite{info15PLB,info16JPG,PPNP18Ma}, which reflects the properties of nuclear matters at finite temperature. With predicted cross sections for proton-rich isotopes, one can also check if the near-proton-drip line isotopes obey this scaling phenomenon.

\section{summary}
\label{summary}
In this work, the cross sections for the proton-rich isotopes in the 140$A$ MeV $^{40}$Ca + $^{9}$Be reaction are predicted by an empirical formula based on the correlation between the cross section and binding energy of fragment. The binding energies in the AME16, WS4 and RCHB methods have been adopted to predict the cross sections for the $Z=$ 10 to 19 isotopes with $T_{z}=-$5 to $-$3/2, and the results are compared to the theoretical results obtained by the {\sc fracs} parametrization and the SAA model. The predicted cross sections for the $T_{z}=-$3 isotopes are close to $10^{-10}$ mb in this work, for which can be hopefully studied in future experiments. The cross sections for the $T_{z} <-$3 isotopes are quite small (smaller than $10^{-14}$ mb), which may be considered in the near future RNB facilities. In addition, based on the cross section distribution and the evident gaps between the $Z=$ 14 nuclei and their neighbors, $Z=$14 is suggested to be a new magic number in the proton-rich nuclei of $T_{z} <-$3/2. And the conclusion is reached that the cross section enhance the signature of new magic numbers in the proton-rich nuclei than that by the average binding energy per nucleon.

\begin{acknowledgments}
This work is partially supported by the National Natural Science Foundation of China (Grant No. U1732135), the Key Research Program of Frontier Sciences of CAS (Grant No. QYZDJSSW-SLH002), the Natural and Science Foundation in Henan Province (Grant No. 162300410179).
\end{acknowledgments}

\end{document}